\documentclass[aps,pre,twocolumn,groupedaddress]{revtex4-2}

\usepackage{amsmath}
\usepackage{bm}
\usepackage{graphicx}
\usepackage{subfigure}
\usepackage{epstopdf}
\usepackage{amssymb}
\usepackage{cases}
\usepackage[colorlinks,
            linkcolor=blue,
            anchorcolor=blue,
            citecolor=blue]{hyperref}

\begin{document}


\title{Electrostatics in a crooked nanochannel in a newly developed curvilinear coordinate system}


\author{Xi Chen, Ke Xiao, Rui Ma, Xuezheng Cao, and Chen-Xu Wu}

\email{cxwu@xmu.edu.cn}
\affiliation{Department of Physics, School of Physical Science and Technology, Xiamen University, Xiamen 361005, People's Republic of China}

\date{\today}

\begin{abstract}
Both biological and artificial nanochannels in crooked shape exhibit unusual transportational characteristics, bringing about a challenge to the traditional theoretical analysis of nanofluidics, partly due to their complicated boundary description. In this paper, by developing a curvilinear coordinate system for crooked nanochannels, we successfully solve the electrostatic Poisson-Boltzmann equation analytically for a two-dimensional nanochannel, with its effectiveness confirmed through numerical calculation. The influences of the geometric profile of the nanochannel on the distribution of electric potential, ionic concentration, and surface charge on channel walls can be quantitatively evaluated in a facilitated way in terms of these curvilinear coordinates. Such a technique can be widely applied to many nanofluidic systems.
\end{abstract}


\maketitle

\section{Introduction}
Nanochannels containing ionic solution in both biological and artificial systems~\cite{duan2010anomalous,hou2011biomimetic,geng2014stochastic} have exhibited many specific features that are absent in macroscopic configurations. In a nanoscale confinement, the interaction between ions and the channel becomes comparatively significant since the suface-to-volume ratio is high. In some biological nanochannels, their wall usually is composed of some functional groups, which will be ionized once in contact with electrolyte solution. Such kind of ionization occurring in the solid-liquid interface makes the channel wall deviate from electroneutrality and greatly influence the ionic distribution in the channel, which, as a result, impacts electrical properties of the whole channel. Many investigations have increasingly been focused on the effect of interfaces between nanochannel walls and electrolyte solution~\cite{hunter2001foundations,schoch2008transport,berg2010introduction}.

In nanoscale, the electrostatic distribution of charged particles near a solid-liquid interface is described as electric double layer (EDL) where charged channel wall attracts counterions to maintain electroneutrality. According to Gouy-Chapman theory~\cite{gouy1910constitution,chapman1913li}, ions in EDL can be regarded as point charges dispersed in uniform dielectric solvent with constant permitivity, so the continuum theory is still valid to describe such nanosystem. In theoretical practice, Poisson-Boltzmann (PB) equation is applied to construct a mean-field model for EDL and its analytical solutions can be expressed in terms of Cartesian coordinates~\cite{hunter1981colloid}, cylindrical coordinates~\cite{rice1965electrokinetic}, or spherical coordinates~\cite{ohshima1982accurate,ohshima2018approximate}, depending on the particular profile of the boundaries.

Previous theoretical investigations on nanochannels were mainly focused on configurations with straight axis, such as cylinder~\cite{vlassiouk2008ionic,vlassiouk2008nanofluidic,movahed2011electrokinetic}, cone~\cite{constantin2007poisson}, or two-dimensional (2D) multilayered-system~\cite{green2015asymmetry} and funnel~\cite{green2018current}, whose axial symmetry can be taken advantage of to simplify governing equations and their corresponding constraint equations in a 2D or even 1D fashion. However, in most cases when such an axial symmetry no longer exists, it becomes extremely hard to analytically solve their governing equations. In Constantin and Siwy's investigation on a conical nanochannel~\cite{constantin2007poisson}, for example, electric potential and ionic concentration were averaged over the transverse coordinates and formulated as functions of axial coordinate. For comparison, they also carried out a numerical simulation for the whole ionic current through the nanochannel  based on the Poisson-Nernst-Planck (PNP) equations without any assumption in advance. When the nanochannel is in an aformensioned regular shape, the analytical solution of PB equation is still attainable via approximation, but in cases of broken axial symmetry, usually one has to resort to numerical simulation instead due to the lack of analytical mathematical tool.

Recent experiments and modelings have discovered some novel phenomena like transportational properties affected by the profile of nanochannels, e.g., curvature plays a significant role in the intercellular exchange between organelles~\cite{rustom2004nanotubular,sherer2007retroviruses} and the electrokinetic flow in artificial nanochannels~\cite{huh2007tuneable}. Unlike straight nanochannels, the distribution of charged particles in a curved channel are governed by curvature-dependent electric potential, which provides a method to dynamically control the directional ion transport inside, i.e., current rectification~\cite{wang2019dynamic,wang2020anomalies}. Meanwhile, in recent nanofluidic experiments, ionic rectification was observed in a system composed of nanosheets fabricated on flexible materials\cite{liu2018two}, which can be treated as a 2D nanochannel with asymmetric geometric configuration, i.e. a 2D curved nanochannel. Such kind of 2D nanochannel can also be found in nanosheets embedded in tunable membranes, a system also showing dynamic shape effect\cite{koltonow2016two,cheng2016ion}.

Although theoretical analyses of curvature effect have been done in rectangular microchannels with uniform curvature~\cite{yun2010geometry,chun2011electrokinetic}, curved nanochannels with varying curvatures along their contours are rarely discussed. Due to the irregularity of its geometric configuration, it is difficult to express the boundary conditions for an electrostatic behavior like EDL based on any traditional coordinate systems, let alone find its analytical solution. Thus it is necessary to develop a suitable coordinate system so as to describe the curved space inside the nanochannel in an easy manner.

In this paper, we develop a curvilinear coordinate system based on Frenet frame, which is found to greatly simplify the representation of the space inside a curved channel and its boundaries. Due to the difficulties in fully solving 3D electrostatic equations analytically, we derive the analytical solution to the PB equation governing the electrostatics inside a 2D curved nanochannel, like the systems reported in Refs.~\cite{liu2018two,koltonow2016two,cheng2016ion}. The effect of channel profile on distribution of electric potential, ionic concentration and surface charge under different boundary conditions is also analyzed. The numerical simulations carried out under the same conditions using finite element method support the effectiveness of our theoretical analysis under such a new curvilinear coordinate frame.

\section{Theoretical analysis of PB equation based on the new curvilinear coordinate system}
\subsection{Curvilinear coordinate system}
\begin{figure}[h]
\flushleft
\subfigure{\label{fig1(a)}
\includegraphics[width=7cm]{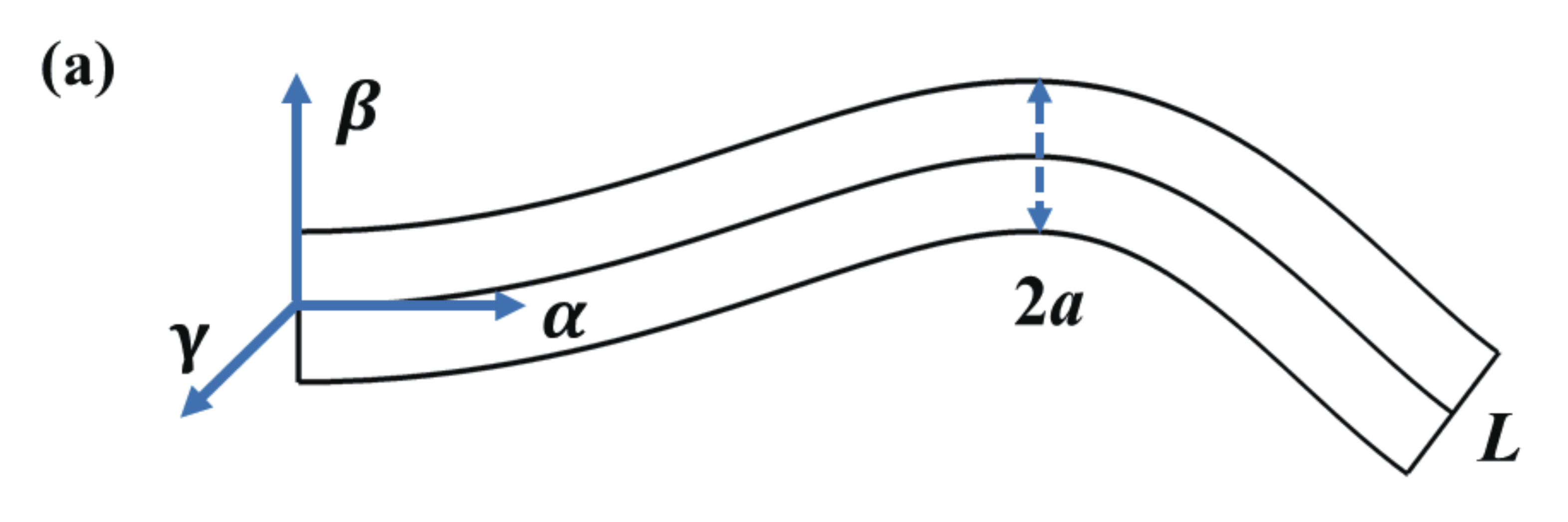}}
\hspace{0in}
\subfigure{\label{fig1(b)}
\includegraphics[width=8.4cm]{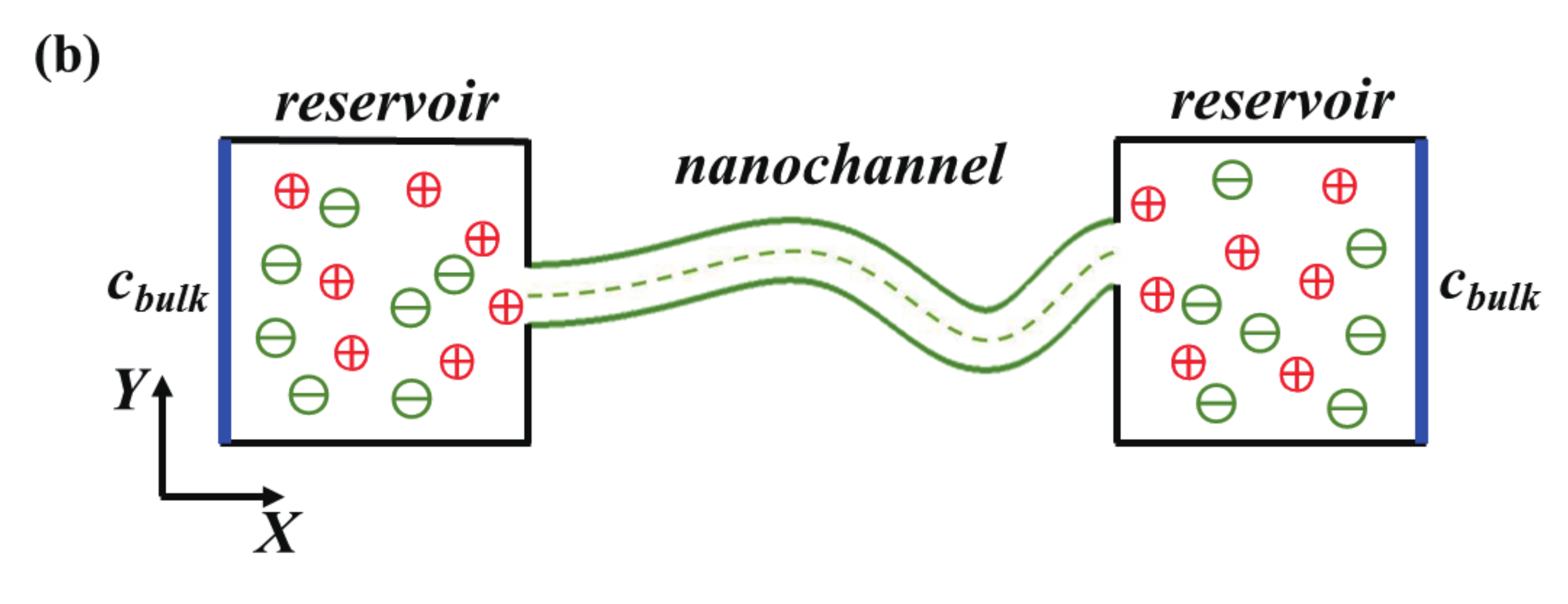}}
\hspace{0in}
\caption{\label{fig1}(a) Curvilinear coordinate system with its origin fixed at the left end of a nanochannel; (b) Schematic picture of a 2D nanochannel connecting two reservoirs of electrolyte solution we used in our numerical simulation. The crooked part in green refers to the nanochannel.}
\end{figure}

In differential geometry theory, Frenet frame is a right-handed orthogonal unit frame moving along a canonical curve with non-zero curvature, which can be defined by Frenet-Serret formula~\cite{struik1961lectures}. Numerous studies have shown its efficacy and convenience in representing vector fields of physical properties inside a curved space, like the electromagnetic field of plasma. For example, for the purpose of analyzing magnetohydrodynamics of the magnetic flux in a toroidal space, like tokamaks~\cite{ricca2005inflexional,garcia2006riemannian}, Frenet frame was utilized to establish an orthogonal curvilinear coordinate frame for the equation of Lorentz force, which depends on the curvature of the tube. Using Frenet frame, the central axis of a planarly crooked nanochannel we consider in $XY$ plane can be expressed, in terms of arc length parameter $s$, as
\begin{equation}
\bm{r_0}(s)=(X(s),Y(s),0).
\end{equation}
The unit tangent vector $\alpha(s)$ at a point $s$ of the axial curve, as shown in Fig. \ref{fig1(a)}, is defined as
\begin{equation}
\bm{\alpha}(s)=(X^{\prime}(s),Y^{\prime}(s),0),
\end{equation}
where the primes refer to the derivatives with respect to the arc length $s$. In this paper, instead of directly following the Frenet frame definition, we directly rotate the tangent vector $\alpha(s)$ counterclockwise for $90^\circ$ in the plane of the axial curve so as to get the unit normal vector at this point, i.e.,
\begin{equation}
\bm{\beta}(s)=(-Y^{\prime}(s),X^{\prime}(s),0),
\end{equation}
and the binormal vector is accordingly defined as
\begin{equation}
\bm{\gamma}(s)=(0,0,1)
\end{equation}
following right-hand rule. Now on the ground of this unit orthogonal frame, we are able to develop a new orthogonal coordinate system in terms of $(s,y,z)$ and thereby the vector function of a specific point $(s,y,z)$ is
\begin{equation}
\bm{r}(s,y,z)=\bm{r_0}(s)+y\bm{\beta}(s)+z\bm{\gamma}(s).
\end{equation}

\subsection{Solution to the PB equation in terms of new curvilinear coordinates}

We consider a 2D crooked nanochannel with charged wall containing binary symmetric electrolyte solution(charge number $z_{\pm}=\pm1$), as shown in Fig.~\ref{fig1}. The nanochannel is connected to two reservoirs with a fixed distance apart as sources of ions (Fig.~\ref{fig1(b)}). In order to study the nanochannel curvature dependence, we let its contour length be variable. According to the Gouy-Chapman's model, counterions of the surface charges are strongly adsorbed to the channel wall, forming a bounded layer of ions, and the concentration of other free counterions tends to decline by the distance from the interface so as to maintain electroneutrality. This screening region with two zones, i.e., the zone of bounded layer and the zone of diffusion layer, is EDL. In electrostatic situations, it is justifiable to consider the ions in EDL to be in quasithermal equilibrium state when no biased potential or extra pressure is applied to the nanochannel, hence the concentration of both kinds of ions is supposed to obey the Boltzmann distribution as a function of the electric potential, i.e.,
\begin{equation}
c_{\pm}=c_{\rm b}\exp\bigg(-\frac{z_{\pm}e\psi}{k_{\rm B}T}\bigg),\label{eq1}
\end{equation}
where $c_b$, $e$, $k_{\rm B}$, and $T$ are the molar bulk concentration far away from the charged channel wall, electric charge, the Boltzmann constant and the temperature (which is set to be 298.15 K in the present calculation) respectively. In addition, on the ground of mean-field assumption, ions in dilute electrolyte solution can be regarded as point charges with their distribution determined by the surrounding electric field, making the electrostatic equation a Poisson-type, which is given by
\begin{equation}
\nabla^2\psi=-\frac{\rho}{\varepsilon_0\varepsilon_{\rm r}},\label{eq2}
\end{equation}
where $\rho = F(c_+-c_-)$, $\varepsilon_0$ and $\varepsilon_{\rm r}$ are the volume charge density, the permittivity of vacuum and the relative permittivity of the electrolyte respectively. Here $F$ is the Faraday constant. Substituting Eq.~(\ref{eq1}) into Eq.~(\ref{eq2}) leads to the PB equation. However, except for some 1D cases, it is extremely hard to solve the PB equation analytically and practically a so-called Debye-H\"{u}ckel (DH) approximation with the premise of low surface electric potential ($\psi_{\rm s}<25$ mV) is frequently used to linearize the PB equation. In DH approximation, expanding the exponential term of electric potential and only keeping the first order term leads to
\begin{equation}
\nabla^2\psi = k^2\psi,
\end{equation}
where $k$, for binary electrolyte, is given by
\begin{equation}
k = \sqrt{\frac{2F^2c_{\rm b}}{\varepsilon_0\varepsilon_{\rm r}RT}}.
\end{equation}
Here $k$ and $R$ are the DH parameter, also defined as the reciprocal of Debye length $\lambda_{\rm D}$, which is the characteristic distance of EDL, and the universal gas constant respectively.

In the curvilinear coordinate system that we develop in the previous section, the scale factors, also called Lam\'{e} coefficients, are defined as
\begin{equation}
h_s=1-yK(s),h_y=1,h_z=1,
\end{equation}
where $K(s)$ stands for the curvature of a planar curve defined in real number field, i.e.,
\begin{equation}
  K(s)^2=X^{\prime\prime}(s)^2+Y^{\prime\prime}(s)^2.
\end{equation}
To be exact, $K(s)$ is positive when the normal vector $\bm{\beta}(s)$ at point $s$ directs toward the center of curvature, and negative when $\bm{\beta}(s)$ points to the opposite direction. With these coefficients, the Laplacian for electric potential can be expressed in terms of these curvilinear coordinates. The 2D crooked nanochannel we consider has a variable contour length $L$ and a width $2a=8$ nm. Its central axis is chosen as an asymmetric one defined by two parametric equations, in terms of $t$, as
\begin{equation}
  \left\{
  \begin{aligned}
  X = & 200t  \\
  Y = & m\sin(\cfrac{5}{2}\pi t^2),
  \end{aligned}
  \right.
\end{equation}
where $m$ is a parameter controlling its profile.
In the 2D nanochannel placed in $XY$ plane, the curvilinear coordinate system can be reduced to a 2D fashion using space variables of arc length $s$ and transverse coordinate $y$. Since typically contour length $L$ of a nanochannel is much larger than its width $2a$, we let $L/a$ be over $50$ so that the curvature $K(s)$ is small if no kinks exist. Therefore it is reasonable to consider only the first-order terms of $yK$. For simplicity and without losing accuracy, the crooked shape of the curved channel can be treated as a deviation from a straight channel, which enables us to apply perturbation theory to solve the PB equation.

First of all we transform the space variables to a dimensionless form in following rules as $\bar{x}=y/a,\bar{s}=s/L,\bar{K}=KL,\bar{k}=ka$ and define a small quantity $\xi=a/L$  for perturbation calculation. Then the PB equation, expanded to the first order of $\xi$, is given by
\begin{equation}
\left(\frac{\partial^2}{\partial \bar{y}^2}-\bar{k}^2\right)\psi+\xi\left(-\bar{y}\bar{K}\frac{\partial^2}{\partial\bar{y}^2}-\bar{K}\frac{\partial}{\partial\bar{y}}+\bar{y}\bar{K}\bar{k}^2\right)\psi=0.
\end{equation}
Correspondingly, we expand the electric potential in terms of $\xi$, keeping the prime and the first-order terms as
\begin{equation}
\psi \approx \psi_0(\bar{y})+\xi\psi_1(\bar{y},\bar{s}).
\end{equation}
Equating the same order terms, we achieve the general solution to the PB equation as
\begin{align}
  &\psi_0(\bar{y}) = c_1e^{\lambda}+c_2e^{-\lambda} \\
  &\psi_1(\bar{y},\bar{s}) = \frac{\bar{K}}{4\bar{k}}[(2c_1\lambda+c_3)e^{\lambda}+(2c_2\lambda+c_4)e^{-\lambda}],
\end{align}
where $\lambda=\bar{k}\bar{y}$ and $c_1, c_2, c_3, c_4$ are all constants to be determined by boundary conditions.

\section{Numerical simulation}
Strictly speaking, the general equations that govern electrokinetic effects in a nanochannel as shown in Fig.~\ref{fig1(b)} are Poisson-Nernst-Planck(PNP) equations~\cite{constantin2007poisson,vlassiouk2008ionic,vlassiouk2008nanofluidic,schoch2008transport}. As the nanochannel reaches the steady state with little convection, the PNP equations combined with continuity equation become
\begin{numcases}{}
  \nabla^2\psi = -\frac{F(c_+-c_-)}{\epsilon} & \label{eq15}\\
  \nabla\cdot\left(D_{\pm}\nabla c_{\pm}+\frac{D_{\pm}Fz_{\pm}c_{\pm}}{RT}\nabla\psi\right) = 0, & \label{eq16}
\end{numcases}
where $D_{\pm}$ is the diffusion coefficients of cations and anions (for simplicity, we assume they are constants\cite{kilic2007steric1,kilic2007steric2}). Note that the PNP equations reduce to the PB equation in thermal equilibrium, a state without flux. The coupled partial differential equations Eqs.~(\ref{eq15}) and~(\ref{eq16}) are numerically solved by using finite element method program COMSOL$^{\rm TM}$. The two insulated reservoirs connected
\begin{figure}[h]
\subfigure{\label{fig2(a)}
\includegraphics[width=8cm]{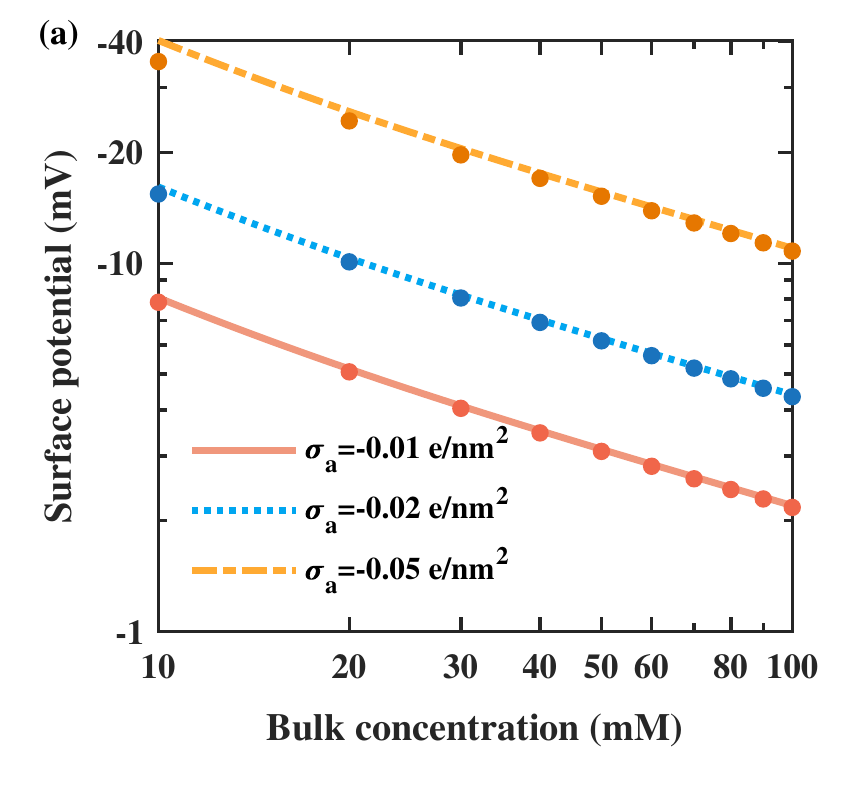}}
\hspace{0in}
\subfigure{\label{fig2(b)}
\includegraphics[width=8cm]{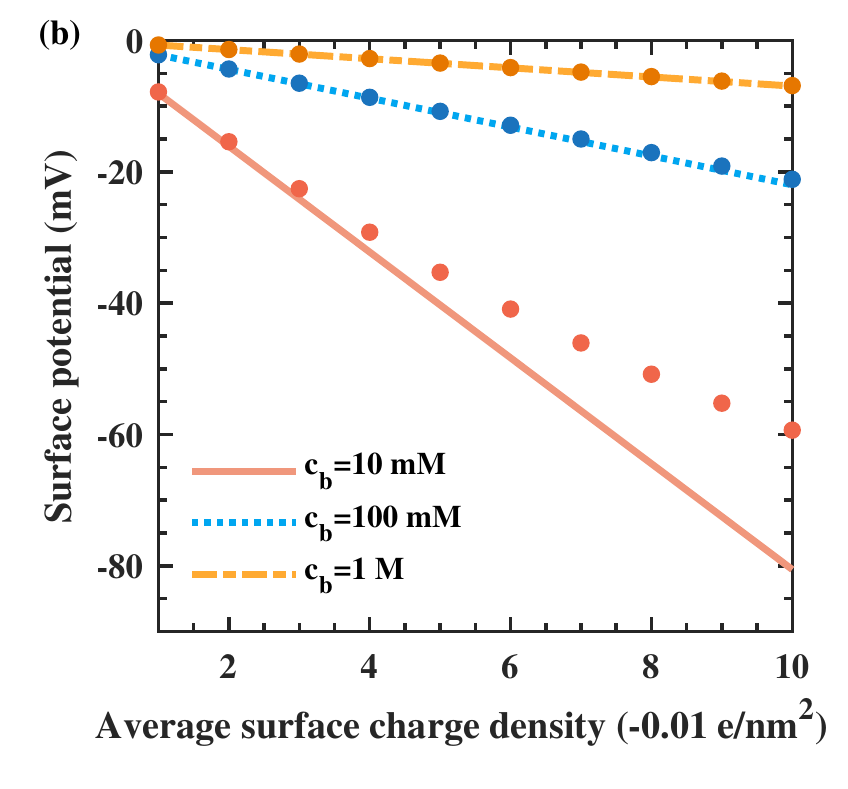}}
\caption{\label{fig2}(a) Surface potential versus bulk concentration ($\psi$-$c_{\rm b}$) for different average surface charge densities ($\sigma_{\rm a}=Q_0/S$). Note that the vertical axis is in reverse order; (b) Surface potential versus average surface charge density ($\psi$-$\sigma$). Line: analytical PB solution; Dot: PNP solution via numerical calculation.}
\end{figure}
to the curved nanochannel, acting as the sources of electrolyte, mathematically play the role of zero electric potential and constant concentration boundary for Eqs.~(\ref{eq15}) and ~(\ref{eq16}). During the process of simulation, the two nanochannel walls are considered as charged surfaces in different boundary conditions, and we fix the distance between the two reservoirs and adjust the profile of the nanochannel via changing the parameter $m$.
\begin{figure}[h]
  \includegraphics[width=8.4cm]{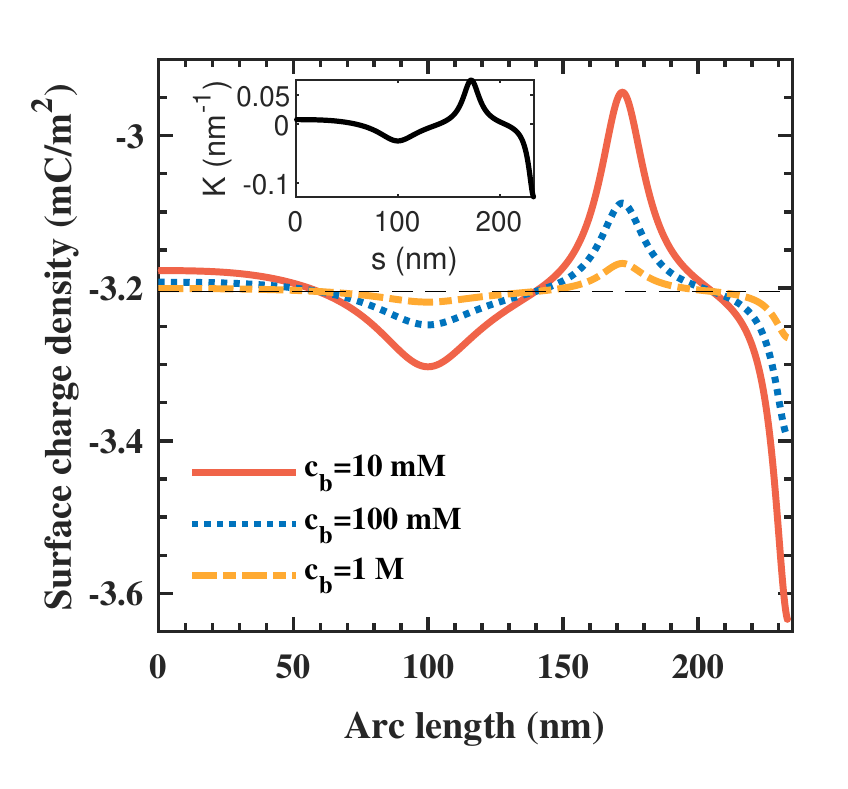}\\
  \caption{\label{fig3}Surface charge density - arc length($\sigma$-s) distribution in different bulk concentrations. Here we only show the surface charge density at the bottom side of the nanochannel because that at the top is nothing but a symmetric one about the black horizontal dashed line ($\sigma=-0.02~{\rm e/nm^2}$). Inset: curvature-arc length (K-s) curve.}
\end{figure}
\section{Results and discussion}
\subsection{Conductive channel walls}
We consider two kinds of boundary conditions for the PB equation, corresponding to different mechanisms of the interaction between nanochannel and electrolyte. Firstly, a conductive boundary, which can be found in lamellar graphene nanochannels~\cite{koltonow2016two,cheng2016ion}, is considered for the 2D nanochannel. An equipotential distribution of the two walls will maintain when they contact with electrolyte solution. More specifically if we consider an asymmetrically crooked nanochannel with each wall carrying a total amount of charge $Q_0$, the boundary condition can be written as
\begin{equation}
\int_S\vec{D}\cdot d\vec{A}=Q_0,
\end{equation}
where $S$, $\vec{D}$ and $\vec{A}$ stand for the surface area, the electric displacement, and the area vector respectively. Figure~\ref{fig2} shows how surface potential of the channel depends on the average surface charge density and concentration of ions in electrolyte solution, where both analytical (lines) and numerical (dots) calculation results are illustrated. A closer look at Fig.~\ref{fig2(a)} exhibits a power law dependence between surface potential and bulk concentration
\begin{equation}
  \psi|_{y=\pm a}\varpropto-c_{\rm b}^{-0.544},
\end{equation}
which is found to decrease linearly with the increase of the total amount of surface charge (Fig.~\ref{fig2(b)}). Such a dependence on bulk concentration comes from the screen effect of the EDL since higher ionic concentration means more ions can be transferred to the surface so as to balance the excess charges.  As expected, most analytical results match well with the numerical simulation results, except for the cases of low bulk concentrations and high surface charge densities.

In order to investigate the electrostatic features along the nanochannel, we also plot the charge density distribution against the arc length, as shown in Fig.~\ref{fig3}. As expected, there exists a tendency of point accumulation at the places of big curvatures, which is found to enhance if the ionic concentration decreases. When the ionic concentration increases, the screen effect of counterions grows stronger and as a consequence the surface charge tends to uniformly distribute along the channel, pushing the distribution curves closer to the black horizontal dashed line, as shown in Fig.~\ref{fig3}. Conclusively, the profile of the nanochannel highly affects the EDL since the Debye length will shorten with the increase of the bulk electrolyte concentration, weakening the boundary effect.
\begin{figure}[h]
\includegraphics[width=8.6cm]{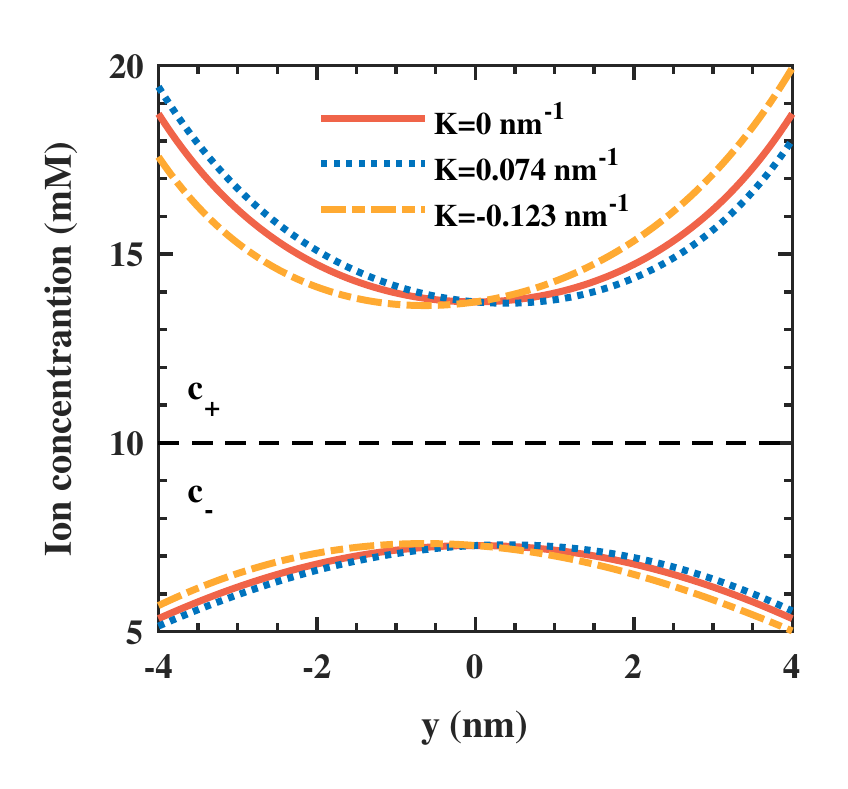}\\
\caption{\label{fig4}Concentration of cations (above) and anions (below) as a function of transverse coordinate($c_{\pm}-y$) at points of different curvatures. Black dashed line indicates the bulk concentration of both kinds of ions.}
\end{figure}
\subsection{Uniformly Charged Channel Walls}
In some biological and artificial systems, the channel wall is coated with functional groups to control transportational properties of a nanochannel, like permeability and permselectivity~\cite{stein2004surface,tunuguntla2017enhanced,kim2007concentration}. For example, nanochannels made of silicate or boron nitride~\cite{siria2013giant} after surface chemical modification will carry immobile
\begin{figure}[h]
\includegraphics[width=8.6cm]{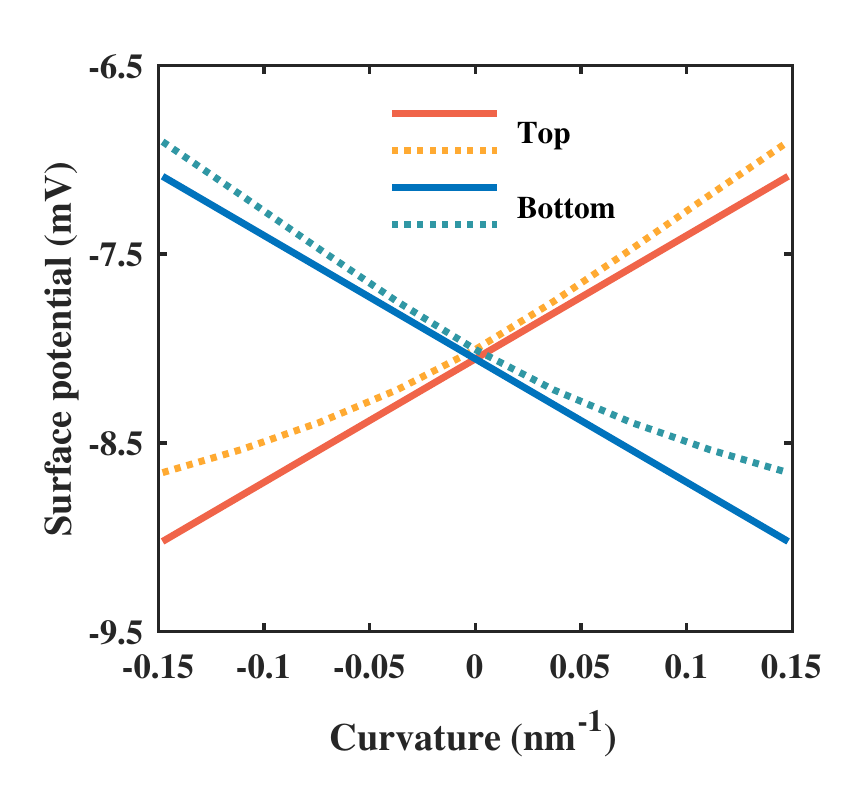}\\
\caption{\label{fig5}Surface potential versus curvature($\psi$-$K$). The surface charge density is set as -0.01 e/nm$^2$ and the bulk concentration is 10 mM. Red and blue lines correspond to top and bottom channel wall respectively. Solid line: PB solution. Dotted line: PNP solution}
\end{figure}
 surface charge in contact with electrolyte solution and theoretically the boundary condition can be considered as fixed surface charge density, like $\sigma_0$, as
\begin{equation}
\left .\mp\varepsilon_0\varepsilon_{\rm r}\frac{\partial\psi}{\partial y}\right|_{\pm a}=\sigma_0.
\end{equation}

Figure~\ref{fig4} shows the analytical result of transverse ionic concentration distribution in the channel at places of different curvatures. As expected, the concentration of cations, decaying from the two channel walls to its axis as a typical feature of EDL, is higher than that of anions in negatively charged nanochannel. Naturally at points of zero curvatures, ionic concentration distribution is symmetric in $y$ direction, while at non-zero curvature points, cations tend to accumulate near the surface that protrudes outward. This preference for staying at concave areas of the channel originates from profile-dependent surface electric potential. Even though the charge density is uniform along the channel wall, the surface electric potential actually varies with the curvature, inducing an asymmetric ionic concentration distribution. Figure~\ref{fig5} presents the linear relationship between surface electric potential and curvature, which provides a possibility of controlling ion distribution in nanochannel via quantitatively adjusting its profile. The reason why the discrepancy between the analytical and the numerical results grow larger as the curvature increases is that high-order terms of curvature in the PB equation is no longer negligible when the curvature radius is reduced to the same scale as the width of the channel.

\section{Conclusion}
We have developed a curvilinear coordinate system for a curved nanochannel, which provides a straightforward and convenient mathematical representation of the static and dynamic process inside. The analytical solutions to the 2D PB equation quantify the profile effect on the electric double layer, which can be seen by the dependence of both surface charge and ionic concentration distributions on the curvature of the nanochannel. These results confirm the possibility to dynamically control the ion distribution as well as the electric potential through bending the nanochannel, a technique that may find its promising application in flexible electronic devices. Further investigation on the curvature effect in 3D nanofluidics can be made after the mathematical challenge of solving the PB equation is overcome. Moreover, a biased potential can be imposed to the two ends of a curved nanochannel so as to induce a directional movement of ions, i.e., an electric current. It has been reported that a rectification on the voltage can be seen in an asymmetrically bent nanochannel~\cite{wang2019dynamic,wang2020anomalies} and the mechanism behind the phenomenon is yet to be clarified.

\begin{acknowledgments}
This work was funded by the National Science Foundation of China under Grant No. 11974292 and No. 11947401.
\end{acknowledgments}

\bibliographystyle{apsrev4-2}

\end{document}